\documentstyle[12pt]{article}
\begin{document}
\phantom{a}

\vspace{1cm}
\begin{center}
{ITP-SB-92-34}
\end{center}
\vspace{1cm}
\begin{center}
{\bf Classification of Superconformal Algebras}
\end{center}
\begin{center}
{\bf with Quadratic Non-Linearity}
\end{center}

\vspace{2cm}
\begin{center}
{\bf E.S.Fradkin and V.Ya.Linetsky}
\end{center}

\vspace{1cm}
\begin{center}
{Department of Theoretical Physics,
P.N.Lebedev Physical Institute, Leninsky pr. 53, Moscow 117924}
\end{center}

\vspace{1cm}
Institute for Theoretical Physics, SUNY at Stony Brook, NY 11794-3840
\begin{center}
{and}
\end{center}
Physics Department, The University of Michigan, Ann Arbor, MI48109
\begin{center}
{\bf Abstract}
\end{center}

A unified treatment of both superconformal and quasisuperconformal algebras
with quad\-ratic non-linearity is given. General formulas describing their
structure are found by solving the Jacobi identities. A complete
classification of quasisuperconformal and ${\bf Z}_2\times{\bf Z}_2$-graded
algebras is obtained and in addition to the previously known cases five
exceptional quasisuperconformal algebras and a series of
${\bf Z}_2\times {\bf Z}_2$-superconformal algebras containing affine
$\widehat{sp}_2\oplus\widehat{osp}(N|2M)$ are constructed.

\newpage
In our previous papers [1] and [2] a complete classification of
superconformal algebras with quadratic non-linearity :JJ: in the
anti-commutator of two supercurrents has been obtained. It is essentially
based on the classification of little finite-dimensional conformal Lie
superalgebras.

Recently another category of non-linear extensions of the Virasoro algebra
has appeared in the literature [3]-[6] in the context of quantum
Hamiltonian reduction, 2D gravity and related matters. These are so-called
quasisuperconformal algebras involving dimension-3/2 bosonic currents in
some representation $\rho$ of a Lie algebra $g$ along with the conventional
stress-energy and $\widehat{g}$ affine Kac-Moody (KM) currents. This category
of
algebras shares many properties with conventional superconformal algebras
obeying the standard spin-statistics relation and, in particular,
quasisuperconformal algebras also can be completely classified.

The goal of the present paper being a sequel of Ref.[1] hereafter referred
to as (I) is to give a unified treatment for both superconformal and
quasisuperconformal algebras and to obtain general formulas determining
their structure. At the beginning we will concentrate on the situation when
$g$ is simple and $\rho$ is irreducible (throughout the main part of this
paper we will be concerned with complex algebras, reality conditions being
discussed in conclusion).

Suppose we are given an irreducible representation $\rho$ of a simple Lie
algebra $g$ with the basis
$\lambda^a=(\lambda^a)_\beta^\alpha
\equiv\lambda^{a,\alpha}_{\phantom{a,}\beta}$,
$[\lambda^a,\lambda^b]=f^{abc}\lambda^c$, $a,b=1,2,\ldots, D:=dim g$,
$\alpha ,\beta,\gamma=1,2,\ldots,d:=dim \rho$.
The basis is chosen so that the structure constants are totally skew. The
Killing metric is then $g^{ab}=-f^{acd}f^{bcd}=-C_v\delta^{ab}$, where
$C_v$ is an eigenvalue of the second Casimir in the adjoint representation
of $g$ related to the dual Coxeter number $h_g^\vee$ by $C_v=\psi^2h_g^\vee$,
where $\psi^2$ is a square of the longest root of $g$. The Dynkin index
$i_\rho$ of the representation $\rho$ is defined according to the
relation $i_\rho=dC_\rho/D\psi^2$, where $C_\rho$ is an eigenvalue of
the second Casimir in the representation $\rho$, i.e.
$\lambda^a\lambda^a=C_\rho {\bf I}$ (and
$tr(\lambda^a\lambda^b)=-i_\rho\psi^2\delta^{ab}$).  (Note that our
definition of the Dynkin index coincides with one in Ref.[7] (in particular
$i_{adg}=h_g^\vee$) and is one half of the definition in Ref.[8].) Greek
indices are raised and lowered with the help of a $g$-invariant metric
$\eta^{\alpha\beta}$ and $\eta_{\alpha\beta}$,
$\eta_{\alpha\gamma}\eta^{\beta\gamma}=\delta^\beta_\alpha$ which is either
symmetric or anti-symmetric
$\eta^{\alpha\beta}=-\varepsilon\eta^{\beta\alpha}$,
depending on either $\rho$ is orthogonal ($\varepsilon=-1$) or
symplectic ($\varepsilon=1$).

Then the most general form of a superconformal ($\varepsilon=-1$) or
quasisuperconformal ($\varepsilon=1$) algebra associated to a given pair
($g$,$\rho$)
reads as follows (note that in this paper we deal only with simple
algebras; the definition of simplicity for non-linear algebras see in (I))

$$
[L_m,L_n] = (m-n)L_{m+n} + \frac{c}{12}m(m^2-1)\delta_{m+n,0},
\eqno{(1a)}
$$

$$
[L_m,J_n^a] = -nJ_{m+n}^a,
\eqno{(1b)}
$$

$$
[J_m^a,J_n^b] = f^{abc}J_{m+n}^c -
\frac{k\psi^2}{2}m\delta^{ab}\delta_{m+n,0},
\eqno{(1c)}
$$

$$
[L_m,G_r^\alpha] = (\frac{m}{2}-r)G_{m+r}^\alpha,
\eqno{(1d)}
$$

$$
[J_m^a,G_r^\alpha] =
-\lambda^{a,\alpha}_{\phantom{a,}\beta}G_{m+r}^\beta,
\eqno{(1e)}
$$

$$
[G_r^\alpha,G_s^\beta]_\varepsilon = 2\eta^{\alpha\beta}L_{r+s} +
\frac{\sigma}{2}(r-s)\lambda^{a,\alpha\beta}J_{r+s}^a$$

$$
+ \frac{b}{2}(r^2-\frac{1}{4})\eta^{\alpha\beta}\delta_{r+s,0} + \gamma
P_{ab}^{\alpha\beta}(J^aJ^b)_{r+s},
\eqno{(1f)}
$$

$$
(J^aJ^b)_m = \sum_{n\in Z}:J_n^aJ_{m-n}^b:,
\eqno{(2)}
$$
where $[A,B]_\varepsilon=AB-\varepsilon BA$ and
$\lambda^{a,\alpha\beta}\stackrel{\hbox{\tiny{def}}}{=}
\lambda^{a,\alpha}_{\phantom{a,}\gamma}\eta^{\beta\gamma}=
\frac{1}{2}(\lambda^{a,\alpha}_{\phantom{a,}\gamma}\eta^{\beta\gamma}+
\varepsilon \lambda^{a,\beta}_{\phantom{a,}\gamma}\eta^{\alpha\gamma})$.
Here $L_m$, $m\in Z$, are Virasoro generators,
$J_m^a$, $m\in Z$,
form an affine KM algebra $\hat{g}$, and
$G_r^\alpha$, $r\in Z(+\frac{1}{2})$,
are dimension 3/2 fermionic (for $\varepsilon=-1$) or bosonic (for
$\varepsilon=1$) currents.  The tensor
$P_{ab}^{\alpha\beta}$ is an interwining operator,
$S^2 adg\stackrel{P}{\longrightarrow}S^2\rho (\Lambda^2\rho$) for
$\varepsilon=-1$ ($\varepsilon=1$), and therefore is fixed by Schur's lemma
up to $p=dim Hom_g$($S^2adg$, $S^2\rho$ (or $\Lambda^2\rho)$)
arbitrary constants. Thus conformal symmetry and $g$-invariance fix the
structure of the algebra (and the corresponding OPE) up to a number of
constants: Virasoro central charge $c$, (conventionally normalized [7]) KM
level $k, b$ (note that our normalization of $b$ in the present paper differs
on one half from the normalization in Refs.[1], [2]), $\sigma$, and $p$
constants inside $P$ (we explicitly display the overall normalization
$\gamma$ of $P$). These quantities are to be fixed by the Jacobi identities
which reduce to the following set of relations among structure constants:

$$\bullet(G,G,G)$$

$$
P_{ab}^{\alpha\beta}\lambda^{b,\gamma}_{\phantom{b,}\delta} +
P_{ab}^{\beta\gamma}\lambda^{b,\alpha}_{\phantom{b,}\delta} +
P_{ab}^{\gamma\alpha}\lambda^{b,\beta}_{\phantom{b,}\delta} = 0,
\eqno{(3a)}
$$

$$
2\eta^{\alpha\beta}\delta_\delta^\gamma -
\eta^{\gamma\alpha}\delta_\delta^\beta -
\eta^{\beta\gamma}\delta_\delta^\alpha -
\frac{\sigma}{2}(\lambda^{a,\beta\gamma}\lambda^{a,\alpha}_{\phantom{a,}\delta}
-\lambda^{a,\gamma\alpha}\lambda^{a,\beta}_{\phantom{a,}\delta})
=-\gamma P_{ab}^{\alpha\beta}
\lambda^{a,\gamma}_{\phantom{a,}\rho}\lambda^{b,\rho}_{\phantom{b,}\delta};
\eqno{(3b)}
$$

$$\bullet(L,G,G)$$

$$
(c-\frac{3}{2}b)\eta^{\alpha\beta} = \frac{\gamma
k\psi^2}{2}P_{aa}^{\alpha\beta};
\eqno{(3c)}
$$

$$\bullet(J,G,G)$$

$$
b = \frac{\sigma k\psi^2}{2},
\eqno{(3d)}
$$

$$
2\eta^{\alpha\beta}\delta^{ab} -
\frac{\sigma}{2}(\lambda^{a,\beta}_{\phantom{a,}\gamma}
\lambda^{b,\alpha\gamma} -
\varepsilon\lambda^{a,\alpha}_{\phantom{a,}\gamma}
\lambda^{b,\beta\gamma}) - \gamma k\psi^2P_{ab}^{\alpha\beta} = \gamma
P_{cd}^{\alpha\beta}f^{ace}f^{bde}.
 \eqno{(3e)}. $$
Note that terms in the right hand sides of Eqs.(3b), (3c) and (3e) are
absent when the Poisson bracket algebra is concerned. They represent
essentially quantum effects of operator normal ordering. Since the
existence of the Poisson bracket algebra is a necessary condition in order
the quantum algebra in question to exist, one finds a necessary condition
for the quantum Jacobi identities to be satisfied

$$
2\eta^{\alpha\beta}\delta_\delta^\gamma -
\eta^{\gamma\alpha}\delta_\delta^\beta -
\eta^{\beta\gamma}\delta_\delta^\alpha -
\frac{\sigma_0}{2}(\lambda^{a,\beta\gamma}
\lambda^{a,\alpha}_{\phantom{a,}\delta} -
\lambda^{a,\gamma\alpha}\lambda^{a,\beta}_{\phantom{a,}\delta})=0,
\eqno{(4)}
$$
for some real $\sigma_0$. Actually, taking the trace in the indices
$\gamma$ and $\delta$, one finds that if the identity (4) holds for the
matrix elements of the representation $\rho$ then

$$
\sigma_0 = -\frac{2\varepsilon(d+\varepsilon)}{C_\rho} =
-\frac{2\varepsilon d(d+\varepsilon)}{\psi^2Di_\rho}.
\eqno{(5)}
$$

At this stage the problem of classification of superconformal
(quasisuperconformal) algebras is reduced to the classification problem for
pairs of simple Lie algebras $g$ and their orthogonal (symplectic)
representations $\rho$ satisfying Eq.(4). For orthogonal representations
$(\varepsilon=-1)$ considered in Ref.[2] it is equivalent to the
classification of little conformal Lie superalgebras with even subalgebras
$sl_2\oplus g$ with simple $g$ (Cases I, V and VI in our classification). For
symplectic $\rho$ $(\varepsilon=1)$ it is equivalent to Cartan's
classification of symmetric subalgebras of the form
${\cal G}_0=sl_2\oplus g$ (with simple $g$) in simple Lie algebras
${\cal G}={\cal G}_0\oplus {\cal G}_1$ (${\cal G}_1$ transforms in the
representation (2,$\rho$) of ${\cal G}_0$ (with irreducible $\rho$)) as
discussed in Note Added to (I) (Cases I and IV-VIII in our list given
therein).

Now one can verify (e.g. by taking advantage of the tables of Ref.[8]) that
in all the cases when (4) does hold one has

$$
S^2\rho = \pi\oplus Id \quad (\Lambda^2\rho = \pi\oplus Id) \qquad
\mbox{for} \quad \varepsilon=-1
(\varepsilon=1) \eqno{(6)} $$
with some irreducible ${\pi}$, and $\pi\oplus Id\in S^2
adg$. Therefore the interwining operator $P$ is defined up
to two constants, one being its overall normalization. Actually it can be
chosen in a form

$$
P_{ab} = \{\lambda_a,\lambda_b\} + 2\nu\delta_{ab}{\bf I}
\eqno{(7)}
$$

($P_{ab}:=(P_{ab})^\alpha_\beta$,
$P_{ab}^{\alpha\beta}=\eta^{\beta\gamma}(P_{ab})^\alpha_\gamma$).
Simple calculation shows that Eq.(3a) is satisfied if and only if

$$
\nu =- \frac{\varepsilon C_\rho}{d+\varepsilon}
\eqno{(8)}
$$
(at this value of $\nu$ it reduces to the identify (4)).

Thus the only free parameters to play with are $k$, $\sigma$ and $\gamma$
($b$ is expressed in terms of $\sigma$ and $k$ by Eq.(3d), and
$c=\frac{1}{2}(3b+\gamma\psi^2\varphi k$) with
$\varphi = 2C_\rho (D+\varepsilon d +1)/(d+2)$ due to
Eq.(3c) and the relation
$P_{aa}^{\alpha\beta}=\varphi\eta^{\alpha\beta}$).Parameters $\sigma$ and
$\gamma$ can be found from the identity (3e). Indeed, taking into account
an identity

$$
f^{ace}f^{bde}P_{cd} = \alpha\{\lambda^a,\lambda^b\} +
\beta\delta^{ab}\bf{I},
\eqno{(9a)}
$$
where
$$
\alpha = C_v - \frac{C_\pi}{2},\quad \beta = 2\nu C_v -
\frac{C_\pi C_\rho}{D},
\eqno{(9b)}
$$
which follows from Eqs.(7), (8) and the branching rule (6)( $C_\pi$ is an
eigenvalue of the second Casimir in $\pi$), it is easy to verify that (3e)
is equivalent to the following two relations among $\sigma$, $\gamma$ and $k$

$$
\varepsilon\sigma + 2\gamma(\psi^2k+\alpha) = 0,
\eqno{(10a)}
$$

$$
2-\gamma(2\psi^2\nu k + \beta) = 0.
\eqno{(10b)}
$$
In this way we are left with the only free parameter $k$.

However, Eq.(3b) imposes one more restriction on $\sigma$, $\gamma$ and $k$

$$
\sigma = \sigma_0 + 2\gamma\chi,\quad \chi := -\frac{d(D+\varepsilon d+1)}
{D}(\frac{C_\rho}{d+\varepsilon})^2,
\eqno{(11)}
$$
which follows from the identity
$$
P_{ab}^{\alpha\beta}\lambda^{a,\gamma}_{\phantom
{a,}\rho}\lambda^{b,\rho}_{\phantom {b,}\delta} =
\chi(2\eta^{\alpha\beta}\delta_\delta^\gamma -
\eta^{\gamma\alpha}\delta_\delta^\beta -
\eta^{\beta\gamma}\delta_\delta^\alpha)
\eqno{(12)}
$$
following in turn from the crucial identity (4). Comparing (10a), (10b) and
(11), we obtain a non-trivial consistency condition

$$
\frac{i_\pi}{i_\rho} = \frac{2d_\pi}{d+\varepsilon},\quad d_\pi := dim\pi =
\frac{d(d-\varepsilon)}{2} - 1
\eqno{(13)}
$$
(where $i_\pi=d_\pi C_\pi/D\psi^2$ is Dynkin's index of
$\pi$). It represents a sufficient condition in order the quantum algebra
(1) to exist. It guarantees that there does not appear any obstruction to
quantization spoiling the quantum Jacobi identities. The condition (13)
seems at first sight quite restrictive. However, as straightforward
verification shows, it does take place in all the cases when the necessary
condition (4) holds. So we find no obstructions to quantization for all the
Poisson bracket superconformal and quasisuperconformal algebras. In
Refs.[1] and [2] we studied two concrete examples, N=7,8 exceptional
algebras associated to $G(3)$ and $F(4)$, and saw that the quantum Jacobi
identities were satisfied owing to various remarkable octonionic
identities. The condition (13) provides a universal criterion for
cancellation of quantum anomalies in the Jacobi identities.

Finally, solving the equations (10a), (10b) and taking into
account Eq.(13), we arrive at the following general expressions

$$
\gamma =\frac{d(d+\varepsilon)}
{\psi^4Di_\rho(k+h_g^\vee+\varepsilon i_\rho)},
\eqno{(14a)}
$$

$$
\sigma = -\frac{2\varepsilon d
[(d+\varepsilon)(k+h_g^\vee)-Di_\rho]}
{Di_\rho(k+h_g^\vee+\varepsilon i_\rho)},
\eqno{(14b)}
$$

$$
b = \frac{k\psi^2\sigma}{2},
\eqno{(14c)}
$$

$$
c = \frac{3}{2}b + \frac{(D+\varepsilon d+1)k}
{k+h_g^\vee+\varepsilon i_\rho}
$$

$$
 = -\frac{3\varepsilon d(d+\varepsilon)k}
{2Di_\rho} + \frac{(D+\varepsilon d+1)(2D+3\varepsilon d)k}
{2D(k+h_g^\vee+\varepsilon i_\rho)}.
\eqno{(14d)}
$$
The above formulas express all the quantities through the KM level
$k$ which remains arbitrary. In Table 1 all the necessary group-theoretic
data are collected which provide the possibility to treat every concrete
case.  Substituting the relevant quantities to Eqs.(14) one convinces
oneself that for $g=so_N,\rho =N$ they coincide with the expressions for
SO(N)-extended superconformal algebras of Refs.[9],[10], for $g=G_2$,
$\rho=7$ - for N=7 exceptional superalgebra of Ref.[1], for g=$Spin(7)$,
$\rho=8_s$ - for exceptional superalgebra of Ref.[2], for $g=sp_{2M}$,
$\rho=2M$ - for the quasisuperconformal algebra of Ref.[5], for last five
exceptional cases it gives five new exceptional quasisuperconformal algebras
the
existence of which has been conjectured in Ref.[2].

Now let us turn to a more general situation when $g$ is reductive and $\rho$
is not necessarily irreducible. The necessary condition for the
quasisuperconformal algebra to exist can now be formulated as a condition
for the existence of a simple Lie algebra ${\cal G}$ possessing a symmetric
subalgebra ${\cal G}_0=sl_2\oplus g$ with the supplementary subspace
${\cal G}_1$ transforming in (2,$\rho$), where $g$ and $\rho$ are no longer
assumed to be simple and irreducible, respectively. Looking at the list of
symmetric subalgebras of simple Lie algebras, one finds two relevant cases
in addition to the six cases with simple $g$ already considered:  ${\cal G}
=sl_{N+2}$, $g=gl_N$, $\rho=N\oplus \bar{N}$ and ${\cal G}
=so_{N+4}$, $g=sl_2\oplus so_N$, $\rho=(2,N)$ (cases II and III in our
list given in Note Added to (I), respectively).  While the
quasisuperconformal algebra associated with $sl_{N+2}$ was constructed in
Refs.[3], [4], the case ${\cal G}=so_{N+4}$ seems to be not explored
previously. We will present the results below in this paper. All the
possibilities for both super and quasisuperconformal algebras with
non-simple $g$ are summarized in Table 2.

In Ref.[6] more general class of algebras has been considered. Therein
dimension-3/2 currents $G_r^\alpha$ carry some representation $\rho$
of a Lie superalgebra $g$, rather than Lie algebra, and dimension-one
currents $J_n^a$ are in the adjoint of $g$. Depending on whether the
indices are even (odd), the corresponding currents are bosonic
(fermionic).  There is a natural ${\bf Z}_2\times{\bf Z}_2$-grading
in the algebras of this category. In Ref.[6] two series of such
${\bf Z}_2\times{\bf Z}_2$-graded superconformal algebras have been
constructed.  They are associated with pairs ($g=osp(N|2M)$, $\rho=N+2M$)
and ($g=sl(N|M)$, $\rho=N+M$) and combine superconformal and
quasisuperconformal algebras together. In Note Added to (I) we conjectured
the existence of a third $\bf{Z}_2\times\bf{Z}_2$-graded series
with $g=sp_2\oplus osp(N|M)$ and $\rho=(2,N+2m)$ which combined the
superconformal series with $g=sp_2\oplus sp_{2M}$ discovered in (I) and
quasisuperconformal series with $g=sp_2\oplus so_N$ [5]. Now we proceed to
describe this item explicitly.

Let ${\cal A,B,C,D}= 1,2,\ldots ,N+2M$ be indices in the fundamental
representation of $osp(N|2M)$. They are handled with the help of an
orthosymplectic metric $B^{\cal AB}$\linebreak $=-(-1)^{P_{\cal A}P_{\cal
B}} B^{\cal BA}$ and $B_{\cal AB}$, $B_{\cal AC}B^{\cal BC}=\delta_{\cal
A}^{\cal B}$, where $P_{\cal A}$ is a Grassmann parity of the index ${\cal
A}$. We assume the first 2M values of ${\cal A}$ denoted by $A$ to be even
($P_A=0$) and the last N denoted by $a$ to be odd ($P_{a}=1$) so that
$B^{\cal AB}$ is block-diagonal:  $B^{AB}:=E^{AB}=-E^{BA}$,
$B^{ab}=-\delta^{ab}$ and $B^{Ab}=B^{bA}=0$. Then our ${\bf Z}_2\times {\bf
Z}_2$-graded superconformal algebra involves dimension-3/2 currents
$G_r^{\alpha {\cal A}}$ ($n(G_r^{\alpha {\cal A}})=P_{\cal A}+1$) and
dimension-one currents $J_m^{\cal{AB}}=(-1)^{P_{\cal A}P_{\cal B}}J_m^{\cal
AB}$ ($n(J_m^{\cal AB})=P_{\cal A}+P_{\cal B}$), along with $L_m$ and
$\widehat{sp_2}$ currents $J_m^{\alpha\beta}=J_m^{\beta\alpha}$
($\alpha,\beta$=1,2). Here $n(\cdot)$ denotes a boson-fermion parity of
currents. It satisfies the following mode algebra (obvious commutators with
$L_m$ and $J_m^{\alpha\beta}$ are omitted)

$$
[J_m^{\cal AB},J_n^{\cal CD}\} = B^{\cal BC}J_{m+n}^{\cal AD} +
B^{\cal AD}J_{m+n}^{\cal BC} + B^{\cal AC}J_{m+n}^{\cal DB} +
B^{\cal BD}J_{m+n}^{\cal CA}
$$

$$
-k_1(B^{\cal AC}B^{\cal BD}-B^{\cal CB}B^{\cal AD})m\delta_{m+n,0},
 \eqno{(15a)} $$

$$
[J_m^{\cal AB},G_r^{\alpha {\cal C}}\} = B^{\cal BC}G_{m+r}^{\alpha
{\cal A}} - B^{\cal CA}G_{m+r}^{\alpha {\cal B}}, \eqno{(15b)} $$

$$
[G_r^{\alpha{\cal A}},G_s^{\beta {\cal B}}] =
2\varepsilon^{\alpha\beta}B^{\cal AB}L_{r+s} +
(r-s)[\sigma_1\varepsilon^{\alpha\beta}J_{r+s}^{\cal AB} +
\sigma_2B^{\cal AB}J_{r+s}^{\alpha\beta}]
$$

$$
+\frac{b}{2}(r^2-\frac{1}{4})\varepsilon^{\alpha\beta}B^{\cal AB}\delta_{r+s,0}
+ \gamma(JJ)_{r+s}^{\alpha A,\beta B}, \eqno{(15c)} $$

$$
(JJ)_m^{\alpha {\cal A},\beta {\cal B}} =
2\varepsilon^{\alpha\beta}B^{\cal AB}
\{\varepsilon_{\alpha_1\beta_1}\varepsilon_{\alpha_2\beta_2}
(J^{\alpha_1\alpha_2}J^{\beta_1\beta_2})_m
- B_{{\cal
A}_1{\cal
B}_1}B_{{\cal
A}_2{\cal B}_2}(J^{{\cal A}_1{\cal A}_2}J^{{\cal B}_1{\cal B}_2})_m\} $$

$$
+ 8(J^{\alpha\beta}J^{\cal AB})_m +
4\varepsilon^{\alpha\beta}B_{\cal CD}\{( J^{\cal AC}J^{\cal BD})_m -
(J^{\cal CB}J^{\cal AD})_m\} \eqno{(16)} $$
(the supercommutators are
defined as usual $[A,B\}=AB-(-1)^{n({\cal A})n({\cal B})}BA$, where
$n(\cdot)$ is a boson-fermion parity defined above). All the constants are
fixed by the super Jacobi identities in terms of one free parameter $k$:

$$
k_1 = -(k+2M-N+4)/2,\quad k_2 = k,
\eqno{(17a)}
$$

$$
\gamma = \frac{1}{2(k-2M+N+4)},
\eqno{(17b)}
$$

$$
b = -4k(k+2M-N+4)\gamma,
\eqno{(17c)}
$$

$$
c = \frac{3}{2}b + [12k+(2M-N+1)(2M-N-4)(k+2M-N+4)]\gamma,
\eqno{(17d)}
$$

$$
\sigma_1 = -4k\gamma,\quad \sigma_2 = 2(k+2M-N+4)\gamma.
\eqno{(17e)}
$$
At N=0 the above formulas coincide with those in (I) describing the
superconformal algebra with $g=sp_2\oplus sp_{2M}$. At M=0 they
define a quasisuperconformal algebra with $g=sp_2\oplus so_N$
promised above.

Above we have considered all the algebras to be complex. Let us now turn to
their real forms. A real form of interest in quantum conformal field theory
is one in which $\hat{g}$ is a compact affine KM algebra. In the basis
chosen in (1c) it is extracted by the standard hermiticity condition

$$
(J_n^a)^\dagger = J_{-n}^a.
\eqno{(18)}
$$
In superconformal algebras ($\varepsilon=-1$) the representation
$\rho$ is orthogonal and consequently one is always able to choose a basis
where $\eta^{\alpha\beta}=\delta^{\alpha\beta}$ and

$$
(G_r^\alpha)^\dagger = G_{-r}^\alpha.
\eqno{(19)}
$$

However in quasisuperconformal algebras ($\varepsilon=1$) $\rho$
symplectic and consequently, for a compact form of $g$, $\rho$ appears to be
pseudoreal, rather than real (in particular, the fundamental representation
of $su_2$ is pseudoreal and it is not possible to define real Euclidean
two-component spinors). The corresponding pseudoreality condition reads as
follows

$$
(G_r^\alpha)^\dagger = i\eta_{\alpha\beta}G_{-r}^\beta
\eqno{(20)}
$$
(one can verify that it preserves the form of the commutation relations
(1)). One has to insert an imaginary unit in Eq.(20) in order to provide
$\dagger^2=Id$. Thus, with the exception of the algebra based on
$sl_{N+2}$, quasisuperconformal algebras possess no real forms with compact
$g$ and real $\rho$ simultaneously, in contrast with usual superconformal
ones. It results in difficulties with unitarity. In case associated with
$sl_{N+2}$ a compact real form does exist owing to the reducibility of
$\rho$. It is based on the real form $su(N+1,1)$ of $sl(N+2;\bf C)$ [3],
[4].

To summarize, the present paper together with Refs.[1] and [2]
constructively solves the classification problem for quantum superconformal
algebras with quadratic non-linearity. We have shown that superconformal
algebras in question exist if and only if there exist corresponding
finite-dimensional structures (little conformal Lie superalgebras for
superconformal algebras with the standard spin-statistics relation, and
$\bf{Z}_2\times\bf{Z}_2$-graded simple Lie algebras with symmetric
subalgebras ${\cal G}_0$ of the form $sl_2\oplus g$ and
${\cal G}_1=(2,\rho)$ for quasisuperconformal algebras; see Tables 1,2 and
3). Meanwhile the second part (only if) of this correspondence is not
unexpected, the first part (if) is quite non-trivial. Indeed we have seen
that the existence of a finite-dimensional algebra is just a necessary
condition. The fact that it also proves to be sufficient (Eq.(13)) seems
quite surprizing. It would be interesting to find out if
there are some deep geometrical reasons for quantum anomalies in the Jacobi
identities to cancel.

In this paper we have been concerned with algebras involving operators with
scale dimensions not higher than two. It is also of interest to study
 algebras involving higher spins and to
obtain complete classification of chiral W-algebras. Our results suggest
that quantum super W-algebras are in one-to-one
correspondence with (perhaps chains of) maximal subalgebras of simple Lie
(super)algebras.\footnote{In this paper we are concerned only with algebras
associated to finite-dimensional structures and parametrized by one
free parameter.At the same time there are a large number
of W-(super)algebras known at present not associated with any
finite-dimensional (super)algebras. However all of them exist only for a
finite set of particular values of the Virasoro central charge and seem to
be certain truncations of larger algebras related with finite-dimensional
algebras (see e.g. Ref.[14] and references therein).} At the classical
level it is in fact realized in the general scheme of the Drinfeld-Sokolov
Hamiltonian reduction [15]. However at the quantum level a problem is
whether any non-linear Poisson bracket W-(super)algebra can be
consistently quantized (i.e. whether there appear no anomalies in the
quantum Jacobi identities). Probably there might exist some generalization
of our anomaly cancellation condition (13) to the more general class of
non-linear W-(super)algebras.

{\bf Acknowledgements}

One of the authors, E.S.Fradkin, would like
to thank C.N.Yang and Warren Siegel for their hospitality.

{\bf Note Added}

After this paper was completed we became aware of a preprint by
P.Bowcock [16] where N=7 exceptional and symplecticaly extended
superconformal algebras of Ref.[1] also were obtained.

\newpage

\noindent
\begin{tabular}{|c|c|c|c|c|c|c|}
\hline
${\cal G}$ &$g$  &$D$  &$h_g^\vee$  &$\rho$
&$\phantom{a}i_\rho\phantom{a}$ &$i_\pi$  $\phantom{\biggr)}$\\
\hline \multicolumn{7}{|c|}
{superconformal ($\varepsilon =-1$)$\phantom{\biggr)}$}\\ \hline
$osp(N|4)$  &$so_N$  &$N(N-1)/2$ &$N-2$ & $N$ &1  &$N+2$
$\phantom{\biggr)}$\\
\hline
$F(4)$  &$so_7$ &21  &5  &$8_s$   &1  &10   $\phantom{\biggr)}$ \\
\hline
$G(3)$   &$G_2$  &14  &4  &7  &1   &9   $\phantom{\biggr)}$\\
\hline
\multicolumn{7}{|c|}
{quasisuperconformal ($\varepsilon =1$)$\phantom{\biggr)}$}\\
\hline
$sp_{2M+2}$ &$sp_{2M}$ &$M(2M+1)$ &$M+1$ &$2M$ &$1/2$ &$M-1$
$\phantom{\biggr)}$\\ \hline $E_6$   &$sl_6$   &35  &6  &20   &3   &54
$\phantom{\biggr)}$\\ \hline
$E_7$    &$so_{12}$  &66  &10   &$32^\prime$   &4  &16
$\phantom{\biggr)}$\\
\hline
$E_8$   &$E_7$    &133  &18  &56  &6   &324   $\phantom{\biggr)}$\\
\hline
$F_4$  &$sp_6$    &21  &4  &$14^\prime$   &$5/2$  &30
$\phantom{\biggr)}$\\ \hline $G_2$  &$sl_2$  &3   &2  &4  &$5/2$  &5
$\phantom{\biggr)}$\\ \hline \end{tabular}

\vspace{0.5cm}
\noindent Table 1.Superconformal and quasisuperconformal algebras with

simple $g$ and irreducible $\rho$.

\newpage

\noindent
\begin{tabular}{|c|c|c|}
\hline
${\cal G}$ &$g$   &$\rho$  $\phantom{\biggr)}$\\
\hline
\multicolumn{3}{|c|}
{superconformal ($\varepsilon =-1$)$\phantom{\biggr)}$}\\
\hline
$sl(2|N)$  &$gl_N$  &$N\oplus\bar{N}$   $\phantom{\biggr)}$\\
\hline
$sl(2|2)/so(2)$  &$sl_2$ &$2\oplus\bar{2}$  $\phantom{\biggr)}$\\
\hline
$osp(4|2M)$  &$sp_2\oplus sp_{2M}$   &$(2, 2M)$ $\phantom{\biggr)}$\\
\hline
$D(1,2;\alpha)$  &$sp_2\oplus sp_2$  &(2,2)     $\phantom{\biggr)}$\\
\hline \multicolumn{3}{|c|}
{quasisuperconformal ($\varepsilon =1$)$\phantom{\biggr)}$}\\ \hline
$sl_{N+2}$  &$gl_N$   &$N\oplus \bar{N}$      $\phantom{\biggr)}$\\
\hline
$so_{N+4}$   &$sp_2\oplus so_N$   &$(2,N)$   $\phantom{\biggr)}$\\
\hline
\end{tabular}

\vspace{0.5cm}
\noindent Table 2.Superconformal and quasisuperconformal algebras
with non-simple $g$ and/or irreducible $\rho$.  \vspace{1cm}

\noindent
\begin{tabular}{|c|c|c|}
\hline
${\cal G}$ &$g$   &$\rho$ $\phantom{\biggr)}$\\
\hline
$osp(N|2M+2)$  &$osp(N|2M)$  &$N|2M$   $\phantom{\biggr)}$\\
\hline
$sl(N|M+2)$  &$sl(N|M)$    &$N|M$   $\phantom{\biggr)}$\\
\hline
$osp(N+4|2M)$  &$sp_2\oplus osp(N|2M)$   &$(2,N|2M)$   $\phantom{\biggr)}$\\
\hline
\end{tabular}

\vspace{0.5cm}
\noindent Table 3.${\bf Z}_2\times{\bf Z}_2$-graded superconformal algebras.
\newpage

\begin{center}
{\bf References}
\end{center}

1. E.S.Fradkin and V.Ya.Linetsky, Zurich preprint ETH-TH/92-6(Feb. 92)

( $Phys.Lett. B$, to be published)

2. E.S.Fradkin and V.Ya.Linetsky, $Phys. Lett.$ {\bf B275} (1992) 345

3. A.M.Polyakov, in "Physics and Mathematics of strings", World
Scientific,1990;

M.Bershadsky, $Commun. Math. Phys.$ {\bf 139} (1991) 71.

4. F.A.Bais, T.Tjin and P. van Driel, $Nucl. Phys.$ {\bf B357}
(1991) 632;

J.Fuchs,preprint CERN-TH.6040/91.

5. L.J.Romans, $Nucl. Phys.$ {\bf B357} (1991) 549;

F.Defever,Ph.D.  Thesis, Leuven 1991.

6. F.Defever, W.Troost and Z.Hasiewicz, $Phys. Lett.$ {\bf B273} (1991) 51.

7. P.Goddard and D.Olive, $Int. J. Mod. Phys.$ {\bf A1} (1986) 303.

8. Slansky, $Phys. Rep.$ {\bf 79} (1981) 1.

9. V.G.Knizhnik, $Theor. Math. Phys.${\bf 66} (1986) 68.

10. M.Bershadsky, $Phys. Lett.$ {\bf B174} (1986) 285.

11. P.Ramond, $Phys. Rev.$ {\bf D3} (1971) 2415;

A.Neveu and J.H.Schwarz, $Nucl. Phys.$ {\bf B31} (1971) 86.

12. M.Ademollo et al., $Phys. Lett.$ {\bf B62} (1976) 105;

$Nucl. Phys.$ {\bf B111} (1976) 77; {\bf B114} (1976) 297;

P.Ramond and J.H.Schwarz, $Phys. Lett.$ {\bf B65} (1976) 75.

13. A.Sevrin, W.Troost and A.Van Proeyen, $Phys. Lett.$ {\bf B208} (1988) 447;

K.Schoutens, $Nucl. Phys.$ {\bf B295} (1988) 634;

P.Goddard and A.Schwimmer, $Phys. Lett.$ {\bf B214} (1988) 209.

14. R.Blumenhagen et al., $Nucl. Phys.$ {\bf B361} (1991) 255.

15. V.I.Drinfeld and V.Sokolov,$J.Sov.Math.${\bf 30} (1984) 1975.

16. P.Bowcock,Preprint Chicago EFI 92-09 (Feb. 92).

\end{document}